\documentclass[12pt,onecolumn,journal,draftclsnofoot,letterpaper]{IEEEtran}
\usepackage{amssymb,amsthm}
\usepackage{thmtools}
\usepackage{amsxtra,amsfonts}
\usepackage[dvipsnames]{xcolor}
\usepackage{enumitem}
\usepackage{mathtools}
\usepackage{enumitem}
\usepackage{latexsym}
\usepackage{mathrsfs}

\newtheorem{lemma}{Lemma}
\newtheorem{theorem}{Theorem}

\usepackage{algorithm}
\usepackage{algorithmic}

\newcommand{\Ind}{\mathcal{I}}
\newcommand{\interior}[1]{%
  {\kern0pt#1}^{\mathrm{o}}%
}

\usepackage{verbatim}
\usepackage{srcltx}

\usepackage[utf8]{inputenc}                                         
\usepackage[T1]{fontenc}
\usepackage[pdfborder={0 0 0},colorlinks,citecolor=blue,hyperindex=true,hypertexnames=false,extension=pdf]{hyperref}

\usepackage[caption=false,font=footnotesize]{subfig}
\usepackage{setspace}

\usepackage{amsmath}
\usepackage{tensor} 

\usepackage{cleveref} 

\newcommand{\Om}{\Omega}

\let\forallalt\forall
\renewcommand{\forall}{\;\forallalt\;}

\let\refalt\ref
\renewcommand{\ref}[1]{(\refalt{#1})}

\newcommand{\alp}{\ensuremath{\alpha}}
\newcommand{\bet}{\ensuremath{\beta}}
\newcommand{\Del}{\ensuremath{\Delta}}

\newcommand{\eps}{\ensuremath{\epsilon}}

\newcommand{\gam}{\ensuremath{\gamma}}

\newcommand{\lam}{\ensuremath{\lambda}}

\newcommand{\Ome}{\ensuremath{\Omega}}
\newcommand{\ome}{\ensuremath{\omega}}

\newcommand{\sig}{\ensuremath{\sigma}}
\newcommand{\tht}{\ensuremath{\theta}}

\newcommand{\tome}{{ \ensuremath{ \tilde{\ome}} }}

\newcommand{\va}{{\ensuremath{\mathbf a}}}
\newcommand{\vb}{{\ensuremath{\mathbf b}}}
\newcommand{\vc}{{\ensuremath{\mathbf c}}}

\newcommand{\vh}{{\ensuremath{\mathbf h}}}                         
\newcommand{\vp}{{\ensuremath{\mathbf p}}}
\newcommand{\vq}{{\ensuremath{\mathbf q}}}

\newcommand{\vx}{{\ensuremath{\mathbf x}}}

\newcommand{\vP}{\ensuremath{\mathbf P }}                         
\newcommand{\vQ}{\ensuremath{\mathbf Q}}

\newcommand{\vome}{\ensuremath{\boldsymbol{ \ome}}}

\newcommand{\tq}{\ensuremath{\tilde{q}}}

\newcommand{\Sset}{\ensuremath{\mathscr S}}

\newcommand{\tvp}{\ensuremath{\tilde{\vp}}}

\newcommand{\R}{{\ensuremath{\mathbb R}}}

\newcommand{\N}{{\ensuremath{\mathbb N}}}

\newcommand{\RA}{\ensuremath{\Rightarrow} }

\newcommand{\LRA}{\ensuremath{\Leftrightarrow} }

\newcommand{\skprod}[2]{\ensuremath{ \left\langle #1,#2 \right\rangle }}

\definecolor{gray}{rgb}{0.3,0.3,0.3}
{\color{black}}                      
{\color{black}}

{\color{black}}                      
{\color{black}}

\newcommand{\thmref}[1]{Theorem~\ref{#1}}     
\newcommand{\lemref}[1]{Lemma~\ref{#1}}       

\newcommand{\appref}[1]{Appendix~\ref{#1}}
\newcommand{\secref}[1]{Section~\ref{#1}}

\newcommand{\figref}[1]{Fig.~\ref{#1}}

\newcommand{\noi}{\noindent}

\newcommand{\Norm}[1]{\ensuremath{ \left\|#1\right\| }}

\newcommand{\Expect}[1]{{\ensuremath{\mathbb E}[#1]}}

\makeatletter
  \newcommand{\set}[2]{\ensuremath{%
  \setbox0=\hbox{\ensuremath{#2}}
  \dimen@\ht0
  \advance\dimen@ by \dp0
  \left\{\left.#1\rule[-\dp0]{0pt}{\dimen@}\;\right|\;#2\right\} }}
\makeatother

\ifx \@paragraph \@empty
\makeatletter
\renewcommand\paragraph{\@startsection
{paragraph}{4}{\z@}{-3.5ex plus-1ex minus-.2ex}%
{1.3ex plus.2ex}{\normalfont\itshape}}
\makeatother
\fi

\DeclareFontFamily{U}{matha}{\hyphenchar\font45}
\DeclareFontShape{U}{matha}{m}{n}{
      <5> <6> <7> <8> <9> <10> gen * matha
      <10.95> matha10 <12> <14.4> <17.28> <20.74> <24.88> matha12
      }{}
\DeclareSymbolFont{matha}{U}{matha}{m}{n}
\DeclareFontSubstitution{U}{matha}{m}{n}

\DeclareFontFamily{U}{mathx}{\hyphenchar\font45}
\DeclareFontShape{U}{mathx}{m}{n}{
      <5> <6> <7> <8> <9> <10>
      <10.95> <12> <14.4> <17.28> <20.74> <24.88>
      mathx10
      }{}
\DeclareSymbolFont{mathx}{U}{mathx}{m}{n}
\DeclareFontSubstitution{U}{mathx}{m}{n}

\DeclareMathDelimiter{\vvvert}{0}{matha}{"7E}{mathx}{"17}

\ifx\bm\undefined
\newcommand{\bm}{\ensuremath{\boldsymbol}}
\fi

\ifx\oast\undefined 
\newcommand{\oast}{\ensuremath{\circledast}}
\fi

\ifdefined\Zero
\renewcommand{\Zero}{\ensuremath{\mathscr Z}}
\else
\newcommand{\Zero}{\ensuremath{\mathscr Z}}
\fi

\ifdefined\Dset
\renewcommand{\Dset}{\ensuremath{\mathfrak D}} 
\else
\newcommand{\Dset}{\ensuremath{\mathfrak D}} 
\fi
\ifdefined\Sset
\renewcommand{\Sset}{\ensuremath{\mathfrak S}} 
\else
\newcommand{\Sset}{\ensuremath{\mathfrak S}} 
\fi

\newif\ifarxiv\arxivfalse
\arxivtrue

\ifarxiv
\usepackage[pdftex]{graphicx}
\fi

\DeclareGraphicsExtensions{.jpg,.pdf,.jpeg,.png,.eps}

\usepackage[style=ieee,    
            url=false,
            backend=bibtex,
            dateabbrev=true,
            doi=false,
            maxbibnames=99,
            isbn=false,
            natbib=true,
            hyperref=true,
            bibencoding=utf8]{biblatex}
\AtEveryBibitem{\clearfield{note}}   
\DefineBibliographyStrings{english}{%
  references = {},
}

\usepackage{svg}
\usepackage{import}

\ifx\remark\undefined
\newenvironment{remark}{\par\vspace{1.5ex}\noindent{\em Remark\/}.}{\par\vspace{1.5ex}}
\fi
\ifx\example\undefined
\newcounter{example}[section]
\newenvironment{example}[1][]{\refstepcounter{example}\par\vspace{1.5ex}\noindent{\em Example~\theexample. #1}}{\par\vspace{1.5ex}}
\fi
\newif\ifproof\prooffalse 

\renewcommand{\vq}{\mathbf p}
\renewcommand{\vp}{\mathbf q}
\renewcommand{\vQ}{\mathbf P}
\renewcommand{\vP}{\mathbf Q}
\newcommand{\Ball}{\ensuremath{\mathcal B}}          
\newcommand{\HS}{\ensuremath{\mathcal{H}}}          
\newcommand{\df}{\ensuremath{\lam}}         
\newcommand{\gP}{\ensuremath{\vP}}          
\newcommand{\gp}{\ensuremath{\vp}}          
\newcommand{\fH}{\ensuremath{\vh}}          
\newcommand{\fh}{\ensuremath{h}}          
\newcommand{\bH}{\ensuremath{\vh}}          
\newcommand{\Pbar}{\ensuremath{\bar{P}}}         
\newcommand{\Rb}{\ensuremath{R_b}}         
\newcommand{\GGT}{\ensuremath{G_{\text{GT}}}}         
\newcommand{\GUAV}{\ensuremath{G_{\text{UAV}}}}         
\newcommand{\Vor}{\ensuremath{\mathcal{V}}}         

\newcommand{\Rset}{\ensuremath{\mathcal R}}

\renewcommand{\Sset}{\ensuremath{\mathcal S}}

\newcommand{\Qset}{\ensuremath{\mathcal P}}

\usepackage{caption}

\newcommand{\Dis}{\ensuremath{D}}                    
\newcommand{\AvDis}{\ensuremath{\bar{\Dis}}}         

\newcommand{\philippstart}{\color{black}}

\makeatletter
\def\blfootnote{\xdef\@thefnmark{}\@footnotetext}
\makeatother 
\begin{document}
%
\title{Quantizers with Parameterized Distortion Measures}

\author{Jun~Guo, Philipp~Walk, and Hamid~Jafarkhani\\
  {\small\begin{minipage}{\linewidth}\begin{center}
    \begin{tabular}{c}
    Center for Pervasive Communications and Computing \\
    University of California, Irvine, CA 92697-2625 \\
    {\it\{guoj4,pwalk,hamidj\}@uci.edu}\\
    \end{tabular}
    \end{center}
  \end{minipage}\vspace{0.2cm}}
}

\maketitle
\thispagestyle{empty}

\begin{abstract}\blfootnote{This work was supported in part by the NSF Award CCF-1815339.}%
  In many quantization problems, the distortion function is given by the Euclidean metric to measure the distance of a
  source sample to any given reproduction point of the quantizer. We will in this work regard distortion functions,
  which are additively and multiplicatively weighted for each reproduction point resulting in a heterogeneous
  quantization problem, as used for example in deployment problems of sensor networks. Whereas, normally in such
  problems, the average distortion is minimized for given weights (parameters), we will optimize the quantization
  problem over all weights, i.e., we tune or control the distortion functions in our favor. 
  For a uniform source distribution in one-dimension, we derive the unique minimizer, given as the uniform scalar
  quantizer with an optimal common weight. By numerical simulations, we demonstrate that this result extends to
  two-dimensions where asymptotically the parameter optimized quantizer is the hexagonal lattice with common weights. As
  an application, we will determine the optimal deployment of unmanned aerial vehicles (UAVs) to provide a wireless
  communication to ground terminals under a minimal communication power cost. Here, the optimal weights relate to the
  optimal flight heights of the UAVs.
\end{abstract}
\section{Introduction}

For a set $\Omega\subset\R^d$ in $d=1,2$ dimensions, a quantizer is given by $N$ reproduction or quantization points
$\vP=\{\vp_1,\dots,\vp_N\}\subset \Ome$ associated with $N$ quantization regions
$\Rset=\{\Rset_1,\dots,\Rset_N\}\subset\Ome$, defining a partition of $\Ome$.
To measure the quality of a given quantizer, the Euclidean distance between the source samples and reproduction points
is commonly used as the distortion function. We will study quantizers with parameter depending distortion functions
which minimize the average distortion over $\Ome$ for a given continuous source sample distribution $\lam:\Ome\to
[0,1]$, as investigated for example in \cite{Erdem16,KJ17,KKSS18} with a fixed set of parameters.
Contrary to a fixed parameter selection, we will assign to each quantization point variable parameters to control the
distortion function of the each quantization point individually.  Such controllable distortion functions widens the
scope of quantization theory and allows one to apply quantization techniques to many parameter dependent network and
locational problems.
In this work, we will consider for the distortion function of $\vp_n$ a Euclidean square-distance, which is
multiplicatively weighted by some $a_n>0$ and additively weighted by some $b_n>0$. Furthermore, we exponentially weight
all distortion functions by some fixed exponent $\gam\geq 1$. To minimize the average distortion, the optimal
quantization regions are known to be generalized Voronoi (Möbius) regions, which can be non-convex and disconnected sets
\cite{BWY07}.
In many applications, as in sensor or vehicle deployments, the optimal weights and parameters are usually unknown, but
adjustable, and one wishes therefore to optimize the deployment over all admissible parameter values, see for example
\cite{ML}.
We will characterize such \emph{quantizers with parameterized distortion measures}  over one-dimensional convex target
regions, i.e., over closed intervals. As a motivation, we will demonstrate such a parameter driven quantizer for an
unmanned aerial vehicle (UAV) deployment to provide energy-efficient communication to ground terminals in a given target
region $\Ome$.  Here, the parameters relate to the UAVs flight heights.  

\paragraph{Notation}  By $[N]=\{1,2,\dots,N\}$ we denote the first $N$ natural numbers, $\N$.  We will
write real numbers in $\R$ by small letters and row vectors by bold letters. The Euclidean norm of $\vx$ is given by
$\Norm{\vx}=\sqrt{\sum_n x_n^2}$.  The open ball in $\R^d$ centered at $\vc\in\R^d$ with radius $r\geq 0$ is denoted by
$\Ball\left(\vc,r\right)=\set{\vome}{\|\vome-\vc\|^2\le r}$.
We denote by $\Vor^c$ the complement of the set $\Vor\subset\R^d$. The positive real numbers are denoted by
$\R_+:=\set{a\in\R}{a>0}$.  Moreover, for two points $\va,\vb\in\R^d$, we denote the generated half space between them,
which contains $\va\in\R^d$, as $\HS(\va, \vb)$.  

\section{System model}\label{sec:model}
To motivate the concept of parameterized distortion measures, we will investigate the deployment of $N$ UAVs positioned at
$\vQ=\{\vq_1,\dots,\vq_N\}\subset(\Ome\times\R_+)^N$ to provide a wireless communication link to ground terminals (GTs) in a
given target region  $\Ome\subset\R^d$. Here, the $n$th UAV's position, $\vq_n=(\vp_n,h_n)$, is given by its ground
position $\vp_n=(x_n,y_n)\in\Ome$, representing the quantization point, and its flight height $h_n$, representing its
distortion parameter.  The optimal UAV deployment is then defined by the minimum average
communication power (distortion) to serve GTs distributed by a density function $\lam$ in $\Ome$ with a
minimum given data rate $R_b$. Hereby, each GT will select the UAV which requires the smallest
communication power, resulting in  so called generalized Voronoi (quantization) regions of $\Ome$,  as used in
\cite{Erdem16,GJ,GJcom18,GJ18,KJ17,ML,MLCS,KKSS18}.  \ifarxiv We also assume that the communication between all users
and UAVs is orthogonal, i.e., separated in frequency or time (slotted protocols).  \fi

\begin{figure}
  \centering
  \def\svgwidth{.85\textwidth} \scriptsize{
    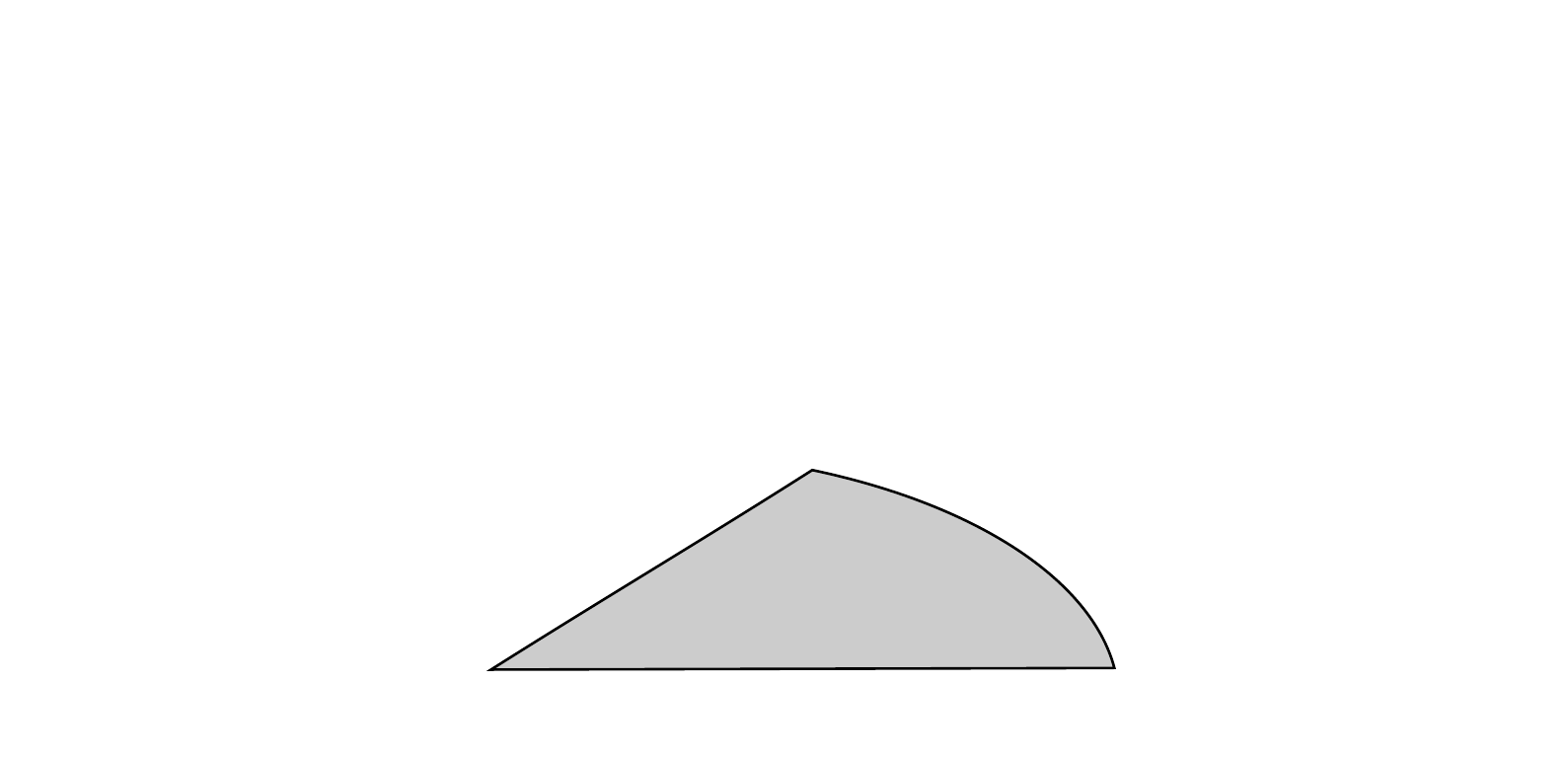}
    \caption{UAV deployment with directed antenna beam and associated GT cells for $\alp=2$ and $N=2$ for a uniform GT distribution.}
    \label{fig:uavdirected}
\end{figure}

In the recent decade, UAVs with directional antennas have been widely studied in the literature
\cite{BJL,MSF,HA,KMR,HSYR,MWMM}, to increase the efficiency of wireless links. However, in
\cite{BJL,MSF,HA,KMR,HSYR,MWMM}, the antenna gain was approximated by a constant within a 3dB beamwidth and set to zero
outside. This ignores the strong angle-dependent gain of directional antennas, notably for low-altitude UAVs. To obtain
a more realistic model, we will consider an antenna gain which depends on the actual radiation angle
$\theta_n\in[0,\frac{\pi}{2}]$ from the $n$th UAV at $\vq_n$ to a GT at $\vome$, see \figref{fig:uavdirected}.
To capture the power falloff versus the line-of-sight distance $d_n$ along with the random attenuation and the
path-loss, we adopt the following propagation model \cite[(2.51)]{Gol05} 
\begin{equation}
  PL_{dB}=10\log_{10}{K}-10\alpha\log_{10}(d_n/d_0)-\psi_{dB},
\end{equation}
where $K$ is a unitless constant depending on the antenna characteristics, $d_0$ is a reference distance, $\alpha\geq 1$
is the terrestrial path-loss exponent, and $\psi_{dB}$ is a Gaussian random variable following
$\mathcal{N}\left(0,\sigma^2_{\psi_{dB}}\right)$. This air-to-ground or terrestrial path-loss model is widely used for
UAV basestations path-loss models \cite{MSBD16a}. Practical values of $\alp$ are between $2$ and $6$ and depend on the
Euclidean distance of GT $\vome$ and UAV $\vq_n$
\begin{align}
 d_n(\vome)= d(\vq_n,(\vome,0))=\sqrt{\|\vp_n-\vome\|^2+h_n^2}=\sqrt{(x_n-x)^2+(y_n-y)^2+h_n^2}\label{eq:eucd}.
\end{align}
For common practical measurements, see for example \cite{AG18}.  Typically maximal heights for UAVs are $<1000$m, due to
flight zone restrictions of aircrafts.  Hence, the received power at UAV $n$ can be represented as
$P_{RX}=P_{TX}G_{TX}G_{RX}Kd^{\alpha}_0 d_n^{-\alpha}(\vome)10^{-\frac{\psi_{dB}}{10}}$,
where $G_{TX}$ and $G_{RX}$ are the antenna gains of the transmitter and the receiver, respectively. Here, we assume
perfect omnidirectional transmitter GT antennas with an isotropic gain and directional receiver UAV antennas.  The angle
dependent antenna gains are 
\begin{equation}
\GGT >0\quad,\quad
  \GUAV = \cos\left(\theta_n\right)=h_n/d_n(\vome),
\label{eq:Gdirected}
\end{equation}
see \cite[p.52]{Bal05a}. The combined antenna intensity is then proportional to
$G=\GUAV \GGT K$, see \figref{fig:uavdirected}.
Accordingly, the received power can be rewritten as
\begin{equation}
  P_{RX}=P_{TX}h_n \GGT K d^{\alpha}_0 d_n^{-\alpha-1}(\vome)10^{-\frac{\psi_{dB}}{10}}.
\end{equation}
To achieve a reliable communication between GT and UAV with bit-rate at least $\Rb$ for a channel bandwidth $B$ and
noise power density $N_0$, the Shannon formula requires $B\log_2\left(1+\frac{P_{RX}}{BN_0}\right)\ge\Rb$.  The minimum
transmission power to UAV $\vq_n$ is then given by%
\ifarxiv
\begin{align}
P_{TX}=\big(2^{\frac{\Rb}{B}}\!-\!1\big)B N_0d(\vq_n,(\vome,0))^{\alpha+1}10^{\frac{\psi_{dB}}{10}} (h_n \GGT
Kd^{\alpha}_0)^{-1}
\end{align}
\fi%
with expectation
\begin{align}
\Expect{P_{TX}}&\! =\!
  \frac{(2^{\frac{\Rb}{B}}-1)N_0 }{ h_n \GGT Kd^{\alpha}_0}
  \frac{d_n^{\alpha+1}(\vome)}{\sqrt{2\pi}\sigma_{\psi_{dB}}} \int_{\R}
     \exp\left(\!-\frac{\psi^2_{dB}}{2\sigma^2_{\psi_{dB}}} \!+\! \ln(10) \frac{\psi_{dB}}{10}\right)\!d\psi_{dB}
     \!=\!\frac{\bet}{h_n}  d_n^{2\gam}(\vome)\label{eq:expectedTX} %
\end{align}
where the independent and fixed parameters are given by
\begin{align}
  \bet=(2^{\frac{\Rb}{B}}-1)B N_0 \exp\big(-\frac{\sig_{\psi_{dB}}^2 (\ln 10)^2}{200}\big)(\GGT K)^{-1}
  d_0^{-\alp}\quad\text{and}\quad
  \gam=\frac{\alp+1}{2}.
\end{align}
Since our goal is to minimize the average transmission power \eqref{eq:expectedTX} we define 
the $n$th \emph{parameter distortion function} as
\begin{align}
  D(\vome,\vp_n,a_n,b_n)=\bet\cdot  \left(a_n\Norm{\vp_n-\ome}_2^2+b_n\right)^\gam
  \label{eq:Eptx}
\end{align}
where $a_n=h_n^{-1/\gam}$ and $b_n=h_n^{2-1/\gam}$.  As can be seen from \eqref{eq:Eptx}, the distortion is a function
of the parameter $h_n$ in addition to the distance between the reproduction point $\vp_n$ and the represented point
$\vome$. From a quantization point of view, one can start with the distortion function \eqref{eq:Eptx} without knowing
the UAV power consumption formulas in this section. This is what we will do in the next section.  For simplicity, we
will set from here on $\bet=1$, since it will not affect the quantizer.

\section{Optimizing Quantizers with parameterized distortion measures}\label{sec:optmize1D}

The communication power cost \eqref{eq:Eptx} defines, with $h_n$ and fixed $\gam\geq 1$, a parameter-dependent
distortion function for $\vp_n$. For a given source sample GT density $\df$ in $\Omega$ and UAV deployment, the average
power is the average distortion for given \emph{quantization and parameter points} $(\vP,\vh)$ with quantization sets
$\Rset=\{\Rset_n\}$, which is called the \emph{average distortion} of the quantizer $(\vP,\vh,\Rset)$
\begin{equation}
 \AvDis(\vP,\vh,\Rset)
  = \sum_{n=1}^N \int_{\Rset_n} D(\vome,\vp_n,h_n)\lam(\vome)d\vome.
  \label{eq:Pbar}
\end{equation}
The $N$ quantization sets, which minimize the average distortion for given quantization and parameter points
$(\vP,\vh)$, define a generalized Voronoi tessellation $\Vor=\{\Vor_n(\vP,\vh)\}$
\begin{align}
  \AvDis(\vP,\vh,\Vor)
  :=\!\int_{\Omega}\min_{n\in[N]} \left\{ \Dis(\vome,\vp_n,h_n) \right\} \df(\vome)d\vome 
  =\sum_{n=1}^{N}\int_{\Vor_n(\vP,\vh)}\!\!\!\!\!\Dis(\vome,\vp_n,h_n) \df(\vome)d\vome 
  \label{eq:optPbar},
\end{align}
where the \emph{generalized Voronoi regions} $\Vor_n(\vP,\vh)$ are defined as the set of sample points $\vome$ with
smallest distortion to the $n$th quantization point $\vp_n$ with parameter $h_n$.  Minimizing the \emph{average
distortion} $\AvDis(\vP,\vh,\Vor)$ over all parameter and quantization points can be seen as an \emph{$N-$facility
locational-parameter optimization problem} \cite{GJ, GJcom18, GJ18,OBSC00}. By the definition of the Voronoi regions
\eqref{eq:optPbar}, this is equivalent to the minimum average distortion over all $N-$level parameter quantizers%
\begin{align}
  \AvDis(\vP^*,\vh^*,\Vor^*)
  = \min_{(\vP,\vh)\in\Ome^N\times\R_+^N} \AvDis(\vP,\vh,\Vor)
  = \min_{(\vP,\vh)\in\Ome^N\times\R_+^N} \min_{\Rset=\{\Rset_n\}\subset\Ome} \AvDis(\vP,\vh,\Rset), 
\label{eq:optquanteqoptdeploy}
\end{align}
which we call the \emph{$N-$level parameter optimized quantizer}.
To find local extrema of \eqref{eq:optPbar} analytically, we will need that the objective function $\AvDis$ be
continuously differentiable at any point in $\Omega^N\times \R_+^N$, i.e., the gradient should exist and be a continuous
function. Such a property was shown to be true for piecewise continuous non-decreasing distortion functions in the
Euclidean metric over $\Ome^N$ \cite[Thm.2.2]{CMB05} and weighted Euclidean metric \cite{GJ}. Then the necessary
condition for a local extremum is the vanishing of the gradient at a critical point\ifarxiv\footnote{Note, if $\nabla
\Pbar$ is not continuous in $\Qset^N$ than any jump-point is a potential critical point and has to be checked
individually.}\fi. 
First, we will derive the generalized Voronoi regions for convex sets $\Om$ in $d$ dimensions for any parameters
$h_n\in\R_+$  for the quantization points $\vp_n$, which are special cases of \emph{M{\"o}bius diagrams
(tessellations)}, introduced in \cite{BWY07}. 
\begin{lemma}\label{lem:moebiusdia}
  Let $\vP=\{\gp_1,\gp_2,\dots, \gp_N\}\subset \Omega^N\subset (\R^d)^N$ for $d\in\{1,2\}$ be the quantization points 
  and $\fH=(\fh_1,\dots,\fh_N)\in\R_+^N$  the associated parameters. Then the average distortion of $(\vP,\vh)$ 
  over all samples in $\Ome$ distributed by $\df$ for  some exponent $\gam\geq1$ 
  \begin{align}
    \AvDis\left(\vP,\fH, \Vor\right) 
    = \sum_{n=1}^{N} \int_{\Vor_n} \! \frac{ (\Norm{\gp_n- \vome}^2 +h_n^2)^{\gam}}{h_n} \df(\vome)d\vome
       \label{eq:minimizationmoebius}
  \end{align}
  has generalized Voronoi regions $\Vor_n= \Vor_{n}(\gP,\fH)= \bigcap_{m\not=n} \Vor_{nm}$, 
  where the dominance regions of quantization point $n$ over $m$ are given by
  \begin{align}
    \Vor_{nm}=\Omega\cap\begin{cases}
         \HS(\gp_n,\gp_m)&, h_m=h_n \\
         \Ball(\vc_{nm},r_{nm}) &, h_n<h_m\\
         \Ball^c(\vc_{nm},r_{nm}) &, h_n>h_m 
        \end{cases}\label{eq:moebius}
  \end{align}
  and center and radii of the balls are given by
  \begin{align}
    \vc_{nm}\!=\!\frac{\gp_n - h_{nm}\gp_m}{1-h_{nm}}
    \quad\text{and}\quad 
    r_{nm}\!=\!\left(\frac{h_{nm}}{\left(1-h_{nm}\right)^2}\Norm{\gp_n-\gp_m}^2  + h_n^2 \frac{h_{nm}^{1-2\gam}
  -1}{1-h_{nm}}\right)^{\frac{1}{2}}.
  \label{eq:rnmcnm}
  \end{align}
  Here, we introduced the parameter ratio of the $n$th and $m$th quantization point as
  \begin{align}
    h_{nm}= \left(h_n/h_m\right)^{\frac{1}{\gam}}.
  \end{align}
\end{lemma}
\ifarxiv
\begin{remark}
  It is also possible that two quantization points are equal, but have different parameters. If the parameter
  ratio is very small or very large, one quantization point can become redundant, i.e., if its optimal quantization set is empty.
  In fact, if we optimize over all quantizer points, such a case will be excluded, which we will show for one-dimension
  in \lemref{lemma:allActive}.
\end{remark}
\fi
\ifarxiv
\begin{proof}
  The minimization of the distortion functions over $\Omega$ defines an assignment rule for a
  generalized Voronoi diagram $\Vor(\gP,\bH)=\{\Vor_1,\Vor_2,\dots,\Vor_N\}$ where 
  \begin{align}
    \Vor_{n} =\Vor_n(\gP,\bH):=
      &\set{\vome\in\Ome}{ a_n\Norm{\vp_n-\vome}^2 +b_n \leq  a_m\Norm{\vp_m-\vome}^2 +b_m, m\not=n}
  \end{align}
  is the $n$th generalized Voronoi region, see for example \cite[Cha.3]{OBSC00}. Here we denoted the weights by the positive numbers
  \begin{align}
    a_n = h_n^{-\frac{1}{\gam}}, \quad b_n=h_n^{2-\frac{1}{\gam}}
  \end{align}
  which define a \emph{M{\"o}bius diagram} \cite{BK06b,BWY07}. The bisectors of M{\"o}bius diagrams are circles or lines
  in $\R^2$ as we will show below.
  The $n$th Voronoi region is defined by $N-1$ inequalities, which  can be written as the intersection of the $N-1$
  \emph{dominance regions} of $\vp_n$ over $\vp_m$, given by 
  \begin{align}
    \Vor_{nm}=\set{\vome\in\Ome}{ a_n\Norm{\vp_n-\vome}^2 +b_n \leq  a_m\Norm{\vp_m-\vome}^2 +b_m}.
  \end{align}
  If $h_n=h_m$ then $a_n=a_m$ and $b_n=b_m$, such that $\Vor_{nm}=\HS(\vp_n,\vp_m)$, the left half-space between $\vp_n$
  and $\vp_m$. For $a_n>a_m$ we can rewrite the inequality as 
  \begin{align*}
      \Norm{\vome}^2 -2 \skprod{\vc_{nm}}{\vome} + 
         \frac{a_n^2 \Norm{\vp_n}^2 \!+\!a_m^2 \Norm{\vp_m}^2 \!-\! a_na_m(\Norm{\vp_n}^2 \!+\!\Norm{\vp_m}^2)}{(a_n-a_m)^2} 
         + \frac{b_n-b_m}{a_n-a_m} \leq & 0
  \end{align*}
  where the center point is given by
  \begin{align}
    \vc_{nm}=\frac{a_n\vp_n- a_m\vp_m}{a_n-a_m}=a_n \frac{\vp_n-h_{nm} \vp_m}{a_n -a_m}=\frac{\vp_n -h_{nm}\vp_m}{1-h_{nm}} 
  \end{align}
  where we introduced the \emph{parameter ratio} of the $n$th and $m$th quantization point as
  \begin{align}
    h_{nm}:= a_m/a_n=\left(h_n /h_m\right)^{\frac{1}{\gam}}>0.
  \end{align}
  If $0<a_n-a_m$, which is equivalent to $h_n<h_m$, then this defines a ball (disc) and for $h_n>h_m$ its complement. Hence we get
  \begin{align}
    \Vor_{nm}=\begin{cases}
      \Ball(\vc_{nm},r_{nm})=\set{\vome\in\Omega}{\Norm{\vome-\vc_{nm}}    <  r_{nm}},&  h_n<h_m\\
      \HS(\vp_n,\vp_m) = \set{\vome\in\Omega}{ \Norm{\vome- \vp_n}\leq \Norm{\vome- \vp_m}}, &h_n=h_m\\
      \Ball^c(\vc_{nm},r_{nm})=\set{\vome\in\Omega}{\Norm{\vome-\vc_{nm}}    >  r_{nm}},&  h_n >h_m
    \end{cases}
  \end{align}
  where the radius square is given by
  \begin{align}
    r_{nm}^2 &= a_na_m\frac{\Norm{\vp_n-\vp_m}^2}{(a_n-a_m)^2} + \frac{b_m-b_n}{a_n-a_m}
              = \frac{a_n}{a_m}\frac{\Norm{\vp_n-\vp_m}^2}{(1-\frac{a_n}{a_m})^2} +
                \frac{b_m-b_n}{a_n-a_m}\label{eq:radiusnm}.
  \end{align}
  The second summand can be written as
  \begin{align}
    \frac{b_m -b_n}{a_n -a_m} &
    = \frac{h_m^{2-\frac{1}{\gam}} - h_n^{2-\frac{1}{\gam}}}{h_n^{-\frac{1}{\gam}}-h_m^{-\frac{1}{\gam}}}
    = \frac{h_n^{2} \left(\left(h_n/h_m\right)^{\frac{1}{\gam}-2} -1\right)}{1-\left(h_n/h_m\right)^{\frac{1}{\gam}}} 
    = h_n^2 \frac{h_{nm}^{-\alp} -1}{1-h_{nm}}
    \label{eq:bmnanm}.
  \end{align}
  For any $\gam\geq 1$, we have $h_{nm}=(h_n/h_m)^{1/\gam}< 1$ if $h_n<h_m$ and $h_{nm}\geq 1$ else.  In
  both cases \eqref{eq:bmnanm} is positive, which implies a radius
  $r_{nm}>0$ whenever $\vp_n\not=\vp_m$. Inserting \eqref{eq:bmnanm} in \eqref{eq:radiusnm} yields the result. 
\end{proof}
\fi 
\begin{example}
We plotted in \figref{fig:uavdirected}, for $N=2$ and $\Omega=[0,1]^2$, the GT regions for a uniform distribution with UAVs
placed on 
\begin{align}
  \vp_1=(0.1 , 0.2), h_1=0.5,\quad\text{and}\quad \vp_2=( 0.6 , 0.6), h_2=1.
\end{align}
If the second UAV reaches an altitude of $h_2\geq 2.3$, its Voronoi region $\Vor_{2}=\Vor_{2,1}$ will be empty and hence
become ``inactive``.
\end{example}

\subsection{Local optimality conditions}
To find the optimal $N-$level parameter quantizer \eqref{eq:optPbar}, we have to minimize the average distortion
\eqref{eq:Pbar} over all possible quantization-parameter points, i.e., we have to solve a non-convex
\emph{$N-$facility locational-parameter optimization problem},
\begin{align}
  \AvDis(\vP^*,\vh^*,\Vor^*)= \min_{\vP\in\Omega^N,\bH\in \R_+^N} \sum_{n=1}^{N} \int_{\Vor_n(\vP,\bH)}
  h_n^{-1}(\Norm{\vp_n-\vome}^2 +h_n^2)^{\gam}\df(\vome)d\vome\label{eq:phoptlocal} 
\end{align}
where $\Vor_n(\vP,\vh)$ are the Möbius regions given in \eqref{eq:moebius} for each fixed $(\vP,\vh)$.
A point $(\vP^*,\vh^*)$ with Möbius diagram $\Vor^*=\Vor(\vP^*,\vh^*)=\{\Vor_1^*,\dots,\Vor^*_N\}$ is a critical
point of \eqref{eq:phoptlocal} if all partial derivatives of $\AvDis$ are vanishing, i.e., if for each $n\in[N]$ it holds
\begin{align}
  0&= \int_{\Vor^*_n} (\vp^*_{n}-\vome) (\Norm{\vp^*_n-\vome}^2 +h_n^{*2})^{\gam-1} \df(\vome)d
  \vome\label{eq:pnopt}\\
  0&= \int_{\Vor^*_n} (\Norm{\vp^*_n-\vome}^2 +h_n^{*2})^{\gam-1}\cdot (\Norm{\vp_n^*-\vome}^2 - (2\gam-1)
  h_n^{*2} ) \df(\vome)d\vome.
  \label{eq:criticalpoint}
\end{align}
For $N=1$ the integral regions will not depend on $\vP$ or $\vh$ and since the integral kernel is continuous differentiable, the
partial derivatives will only apply to the integral kernel. For $N>1$, the conservation-of-mass law, can be used to show
that the derivatives of the integral domains will cancel each other out, see also \cite{CMB05}. 
\ifarxiv
\begin{remark}
  The shape of the regions depend on the parameters, which if different for each quantization point (heterogeneous),
  generate spherical and not polyhedral regions. We will show later, that homogeneous parameter selection with polyhedral
  regions will be the optimal regions for $d=1$. 
\end{remark}
\fi

\subsection{The optimal $N-$level parameter quantizer in one-dimension for uniform density}

In this section, we discuss the parameter optimized quantizer for a one-dimensional convex source $\Ome\subset\R$, i.e.,
for an interval $\Omega = [s,t]$ given by some real numbers $s<t$.  Under such circumstances, the quantization points
are degenerated to scalars, i.e., $\vp_n=x_n\in[s, t], \forall n\in[N]$. If we shift the interval $\Ome$ by an arbitrary
$a\in\R$, then the average distortion, i.e., the objective function, will not change if we shift all quantization points
by the same number $a$. Hence, if we set $a=-s$, we can shift any quantizer for $[s,t]$ to $[0,A]$ where $A=t-s$ without
loss of generality.
Let us assume a uniform distribution on $\Omega$, i.e.  $\df(\ome)=1/A$. 
To derive the unique $N-$level parameter optimized quantizer for any $N$, we will first investigate the case $N=1$.
\begin{lemma}\label{lem:ggam}
  Let $A>0$ and $\gam\geq 1$. The unique $1-$level parameter optimized quantizer $(x^*,h^*)$ with distortion
  function \eqref{eq:Eptx} is given  for a uniform source
  density in $[0,A]$ by
  \begin{align}
    x^*\!=\!\frac{A}{2},   h^*\!=\!\frac{A}{2} g(\gam) \quad\text{and the minimum average distortion} \quad
    \AvDis(x^*,h^*)\!=\!\left(\frac{A}{2}\right)^{\!2\gam-1}\!\!\!\!g(\gam)\notag
  \end{align}
  where  $g(\gam)=\arg\min_{u>0} F(u,\gam)<1/\sqrt{2\gam-1}$ is the unique minimizer of 
  \begin{align}
    F(u,\gam)=\int_0^1  f(\ome,u,\gam) d\ome \quad\text{with}\quad f(\ome,u,\gam)=\frac{(\ome^2+u^2)^{\gam}}{u} 
  \end{align}
  which is for fixed $\gam$ a continuous and convex function over $\R_+$. For $\gam\in\{1,2,3\}$ the minimizer can be
  derived in closed form as 
  \begin{align}
    g(1) = \sqrt{1/3},\ \  g(2) = \sqrt{ (\sqrt{32/5}-1)/9}, 
    \ \  g(3) = \sqrt{\Big((32/7)^{1/3}-1\Big)/5}.\label{eq:ggam}
  \end{align}
\end{lemma}
\ifarxiv 
\begin{proof}
See \appref{app:proof_lemma_ggam}.
\end{proof}
\fi

\begin{remark}
  The convexity of $F(\cdot,\gam)$ can be also shown by using extensions of the Hermite-Hadamard inequality \cite{ZC10},
  which allows to show convexity over any interval.
  Let us note here that for any fixed parameter $h_n>0$, the average distortion $\AvDis(x_n^*\pm \eps,h_n)$ is strictly
  monotone increasing in $\eps>0$. Hence, $x_n^*$ is the unique minimizer for any $h_n>0$. We will use this decoupling
  property repeatedly in the proofs \cite{GWJ18b}.
\end{remark}

\newcommand{\pGlob}{\ensuremath{x^{*}}}
\newcommand{\hGlob}{\ensuremath{h^{*}}}
\newcommand{\qGlob}{\ensuremath{{\mathbf q}^{*}}}
\newcommand{\qLoc}{\ensuremath{{\mathbf q}^{*}}}
\philippstart
\color{black}
To derive our main result, we need some general properties of the optimal regions. 
\philippstart
\begin{lemma}\label{lemma:allActive}
  Let $\Ome=[0,A]$ for some $A>0$. The $N-$level parameter optimized quantizer $(\vP^*,\vh^*)\in\Ome^N\times\R_+^N$ for
  a uniform source density in $\Ome$ has optimal quantization regions $\Vor_n(\vP^*,\vh^*)=[b^*_{n-1},b^*_n]$ with
  $0\leq b^*_{n-1}<b^*_n\leq A$ and optimal quantization points $x_n^*=(b_n^*+b^*_{n-1})/2$ for $n\in[N]$, i.e., each
  region is a closed interval with positive measure and centroidal quantization points.
\end{lemma}
\ifarxiv
\begin{proof}
  See \appref{app:proof_lemma_active}.
\end{proof}
\fi
\begin{remark}
  Hence, for an $N-$level parameter optimized quantizer, all quantization points are used, which is intuitively, since
  each additional quantization point should reduce the distortion of the quantizer by partitioning the source in non-zero regions. 
\end{remark}
%
%
\begin{theorem}\label{thm:commonheight}
  Let $N\in\N$,  $\Ome=[0,A]$ for some $A>0$, and $\gam\geq 1$. The \emph{unique $N-$level parameter optimized quantizer} 
  $(\vP^*,\vh^*,\Rset^*)$ is the  uniform scalar quantizer with identical parameter values, given for $n\in[N]$  by   
  \begin{align}
    \vp_n^*=\pGlob_n= \frac{A}{2N} (2n-1),\quad   h^*=h_n^*= \frac{A}{2N} g(\gam),\quad \Rset_n^*=
    \left[\frac{A}{N}(n-1),\frac{A}{N}n\right]
    \label{eq:optimaldeploy} 
  \end{align}
  with minimum average distortion
  \begin{align}
   \AvDis(\vP^*,\vh^*,\Rset^*)=\left(\frac{A}{2N}\right)^{2\gam-1}  \int_0^1 \frac{\big(\ome^2+g^2(\gam)\big)^{\gam}}{g(\gam)}
    d\ome\label{eq:optimumavpow}.
  \end{align}
  For $\gam\in\{1,2,3\}$, the closed form $g(\gam)$ is provided in  \eqref{eq:ggam}.
\end{theorem}

\ifarxiv
\begin{proof} See \appref{sec:proof_theorem}.
\end{proof}
\fi 
\color{black}

\begin{figure}
\begin{minipage}[b]{0.4\textwidth}
  \vspace{1ex}
\def\svgwidth{1.1\textwidth} \scriptsize{
  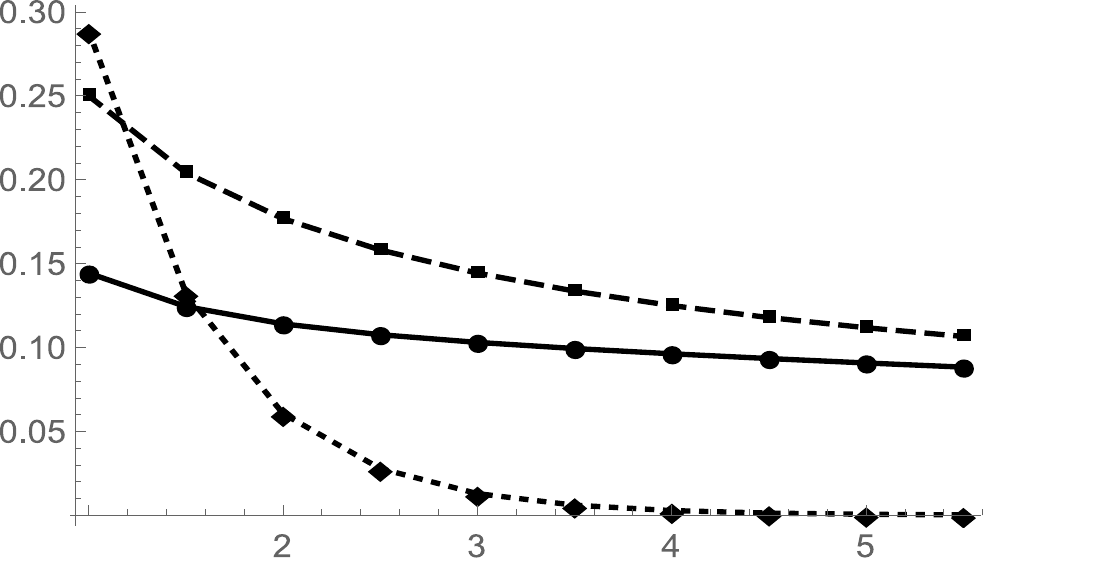}
  \vspace{-3.2ex}
  \caption{{\small Optimal height (solid) with bound (dashed) and average distortion (dotted) for 
  $N=2,A=1$ and uniform GT density.}}
  \label{fig:goptdopt}
\end{minipage}
\hfill
  \begin{minipage}[b]{0.56\textwidth}
\hspace{-2ex} 
    \def\svgwidth{1.08\textwidth} \scriptsize{
      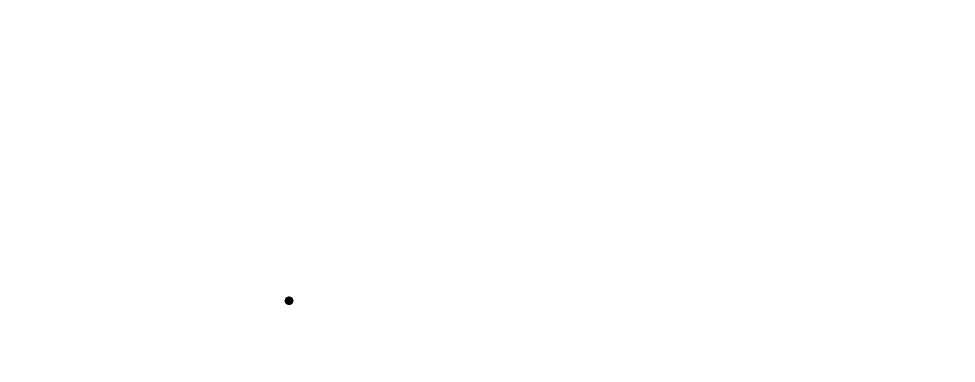}
      \caption{{\small Optimal UAV deployment in one dimension for $A=1,\alp=1$ and $N=2,4$ over a uniform GT density
      by \eqref{eq:optimaldeploy}.}}
      \label{fig:uavonedim}
    \end{minipage}
\end{figure}

\begin{example}
  We plot  the optimal heights and optimal average distortion for a uniform GT density in $[0,1]$ over various $\alp$
  and $N=2$ in \figref{fig:goptdopt}. Note that the factor $A/2N=1/4$ will play a crucial role for the height and
  distortion scaling. Moreover, the distortion decreases exponentially in $\alpha$ if $A/2N<1$.

  Let us set $\bet=1=A$. Then, the optimal UAV deployment is pictured in \figref{fig:uavonedim} for $N=2$ and $N=4$. The
  maximum elevation angle $\tht_{\text{max}}$ is hereby constant for each UAV and does not change if the number of UAVs,
  $N$, increases.  Moreover, it is also independent of $A$ and $\bet$, since with \eqref{eq:optimaldeploy} we have
  $\mu_n^*=x^*_n-x^*_{n-1}=A/N$ and
  \begin{align}
    \cos(\tht_{\text{max}})=\cos(\tht_n)= \frac{h^*}{\mu^*_n/2}= \frac{2N}{A}\frac{A}{2N}   g(1)= \frac{1}{\sqrt{3}}.
  \end{align}
\end{example}

\section{Llyod-like Algorithms and Simulation Results}
In this section, we introduce two Lloyd-like algorithms, Lloyd-A and Lloyd-B, to optimize the quantizer for
two-dimensional scenarios. The proposed
algorithms iterate between two steps: (1) The reproduction points are optimized through gradient descent while the
partitioning is fixed; (ii) The partitioning is optimized while the reproduction points are fixed.  In Lloyd-A, all UAVs
(or reproduction points) share the common flight height while Lloyd-B allows UAVs at different flight heights.

In what follows, we provide the simulation results over the two-dimensional target region $\Omega=[0,10]^2$ with
uniform and non-uniform density functions.  The non-uniform density function is a Gaussian mixture of the form
$\sum_{k=1}^{3}\frac{A_k}{\sqrt{2\pi}\sigma^2_k}\exp{\left(-\frac{\|\ome-c_k\|^2}{2\sigma_k}\right)}$, where the
weights, $A_k$, $k=1,2,3$ are $0.5$, $0.25$, $0.25$, the means, $c_k$, are $(3,3)$, $(6,7)$, $(7.5,2.5)$, the standard deviations,
$\sigma_k$, are  $1.5$, $1$, and $2$, respectively.

\begin{figure}[!htb]
\setlength\abovecaptionskip{0pt}
\setlength\belowcaptionskip{0pt}
\centering
\subfloat[]{\includegraphics[width=2.9in]{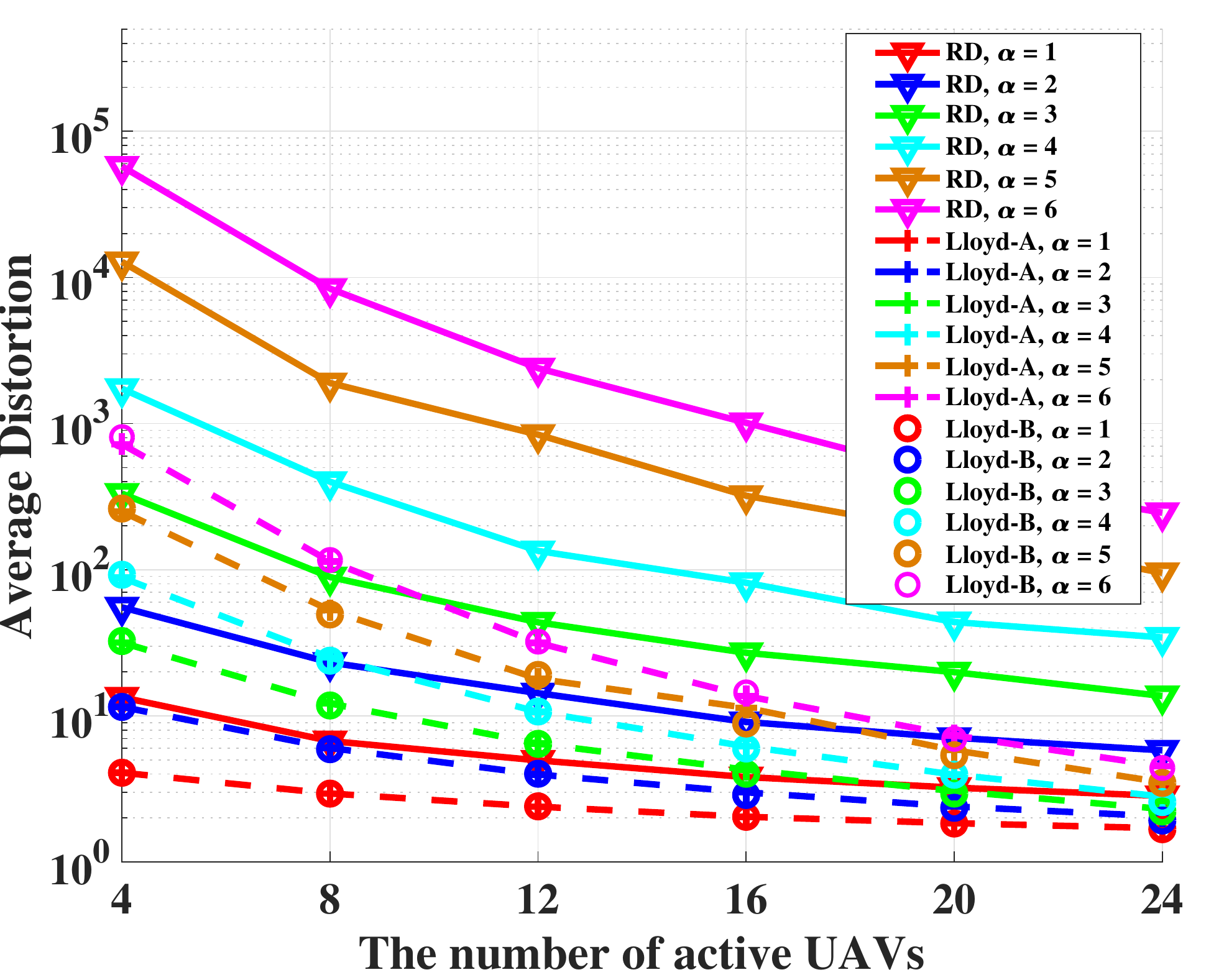}
\label{uniformDistortion}}
\hfil
\subfloat[]{\includegraphics[width=2.9in]{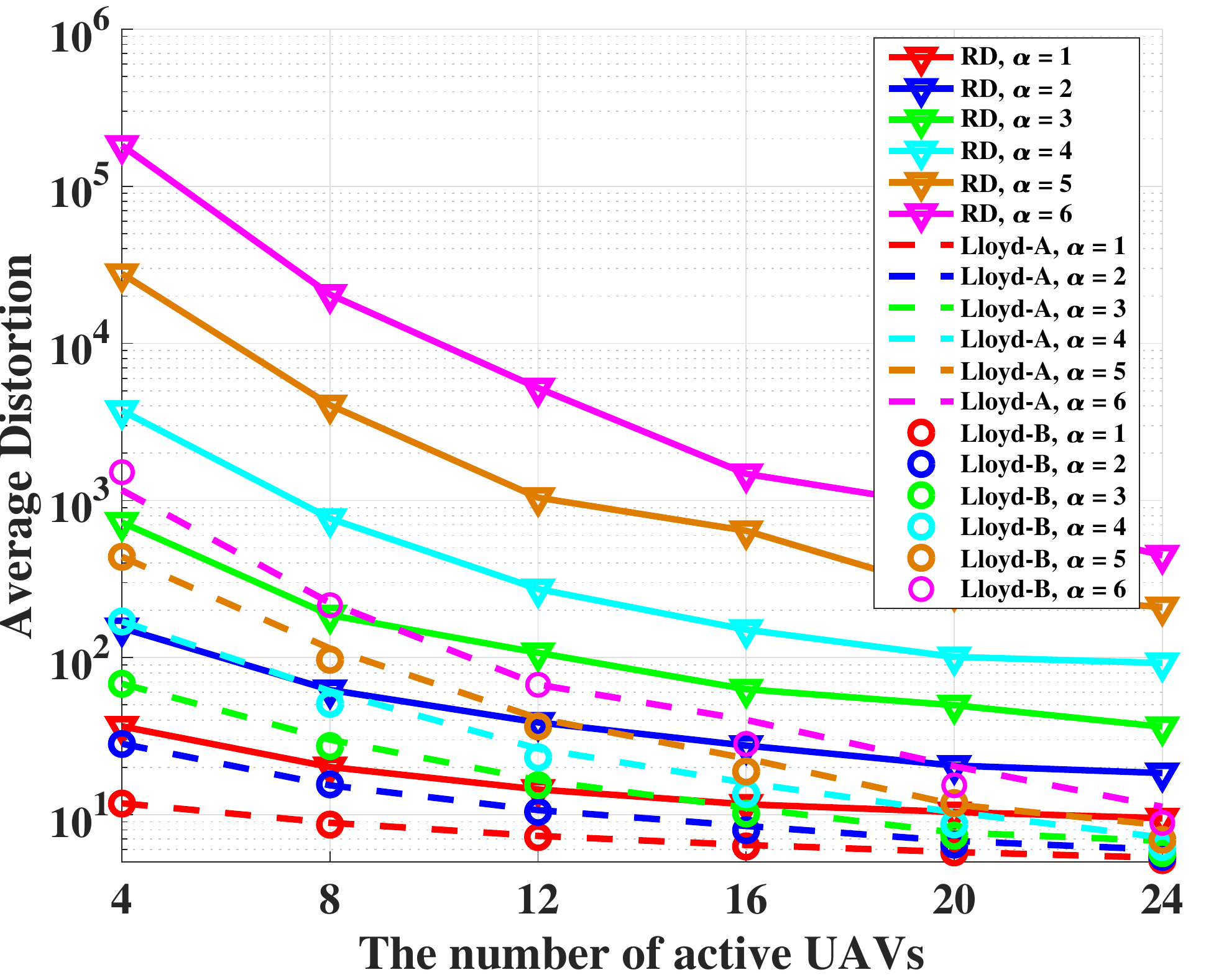}
\label{nonuniformDistortion}}
\captionsetup{justification=justified}
\caption{\small{The performance comparison of Lloyd-A, Lloyd-B and Random Deployment (RD). 
(a) Uniform density. (b) Non-uniform density.}}
\label{Distortion}
\end{figure}

\begin{figure}[t]
\centering
\subfloat[]{\includegraphics[width=2.8in]{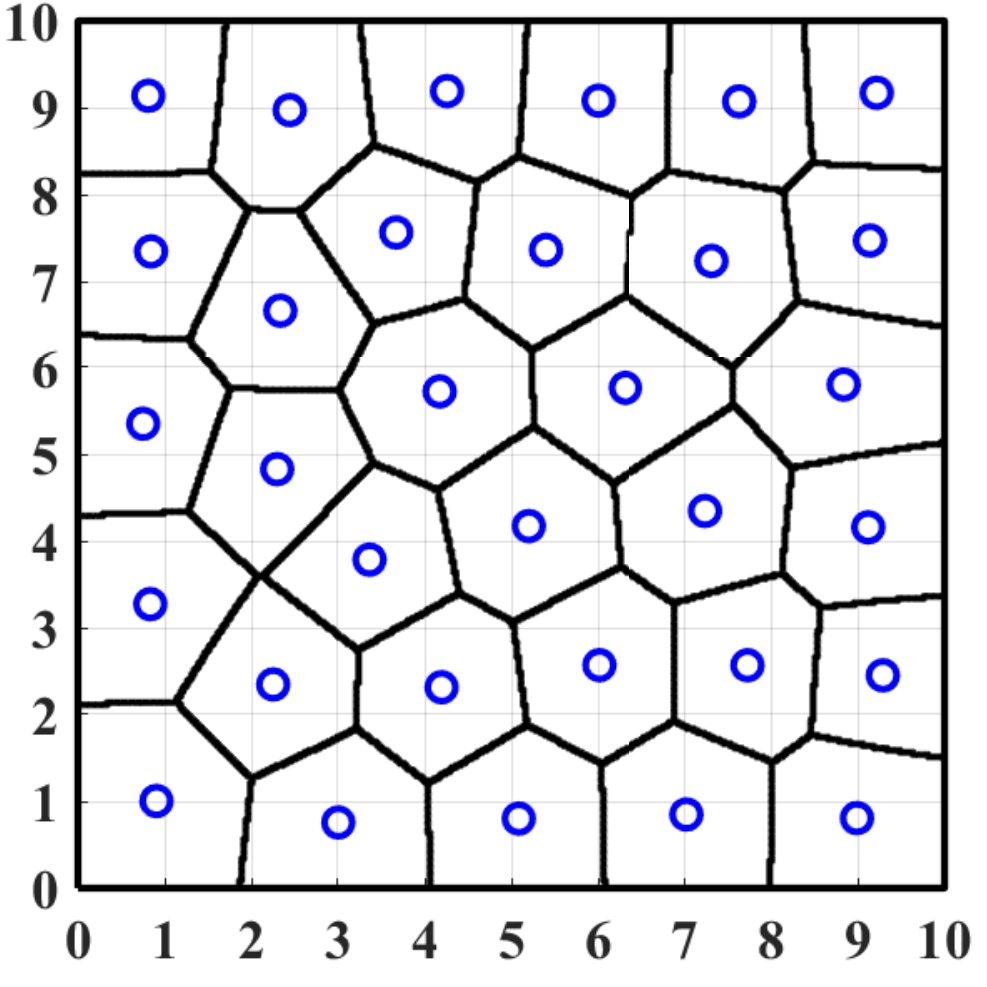}
\label{uniformPartitions32}}
\hfil
\subfloat[]{\includegraphics[width=2.8in]{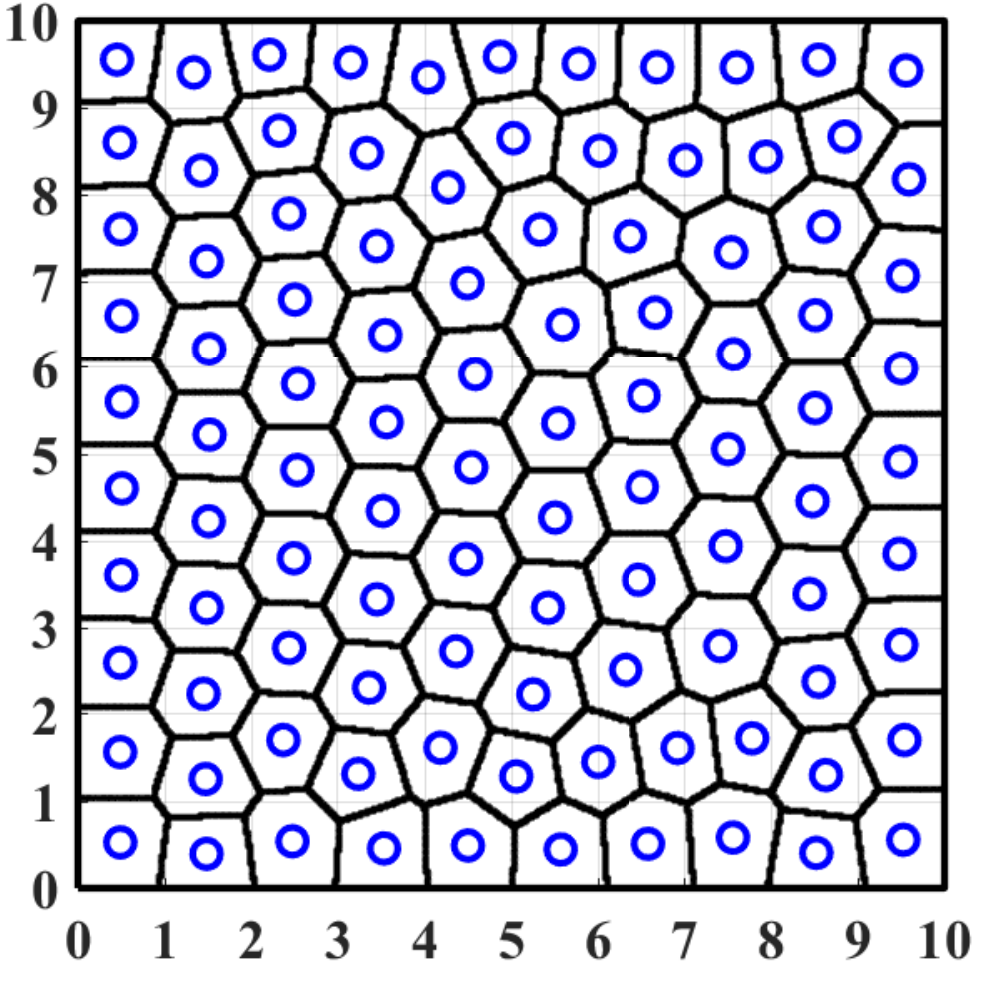}
\label{uniformPartitions100}}
\captionsetup{justification=justified}
\vspace{-2ex}
\caption{\small{The UAV projections on the ground with generalized Voronoi Diagrams where $\alpha=2$ and the source distribution is uniform. 
(a) 32 UAVs. (b) 100 UAVs.}}
\label{uniformDistortionPartition2}
\end{figure}

To evaluate the performance of the proposed algorithms, we compare them with the average distortion of $100$ random
deployments (RDs).  Figs. \ref{uniformDistortion} and \ref{nonuniformDistortion}, show that the proposed algorithms
outperform the random deployment on both uniform and non-uniform distributed target regions.  From
\figref{uniformDistortion}, one can also find that the distortion achieved by Lloyd-A and Lloyd-B are very close,
indicating that the optimality of the common height, as proved for the one-dimensional case in \secref{sec:optmize1D},
might be extended to the two-dimensional case when the density function is uniform. However, one can find a
non-negligible gap between Lloyd-A and Lloyd-B in \figref{nonuniformDistortion} where the density function is
non-uniform. For instance, given $16$ UAVs and the path-loss exponent $\alpha=6$, Lloyd-A's distortion is $40.17$ while
Lloyd-B obtains a smaller distortion, $28.25$, by placing UAVs at different heights.  
Figs. \ref{uniformPartitions32} and \ref{uniformPartitions100} illustrate the UAV ground projections and their
partitions on a uniform distributed square region. As the number of UAVs increases, the UAV partitions approximate
hexagons which implies that the optimality of congruent partition (Theorem \ref{thm:commonheight}) might be extended to
uniformly distributed users for two-dimensional sources.  \ifarxiv However, the UAV projections in Figs.
\ref{nonuniformPartitions32} and \ref{nonuniformPartitions100} show that congruent partition is no longer a necessary
condition for the optimal quantizer when distribution is non-uniform. \fi
\ifarxiv
\begin{figure}[t]
\centering
\subfloat[]{\includegraphics[width=2.8in]{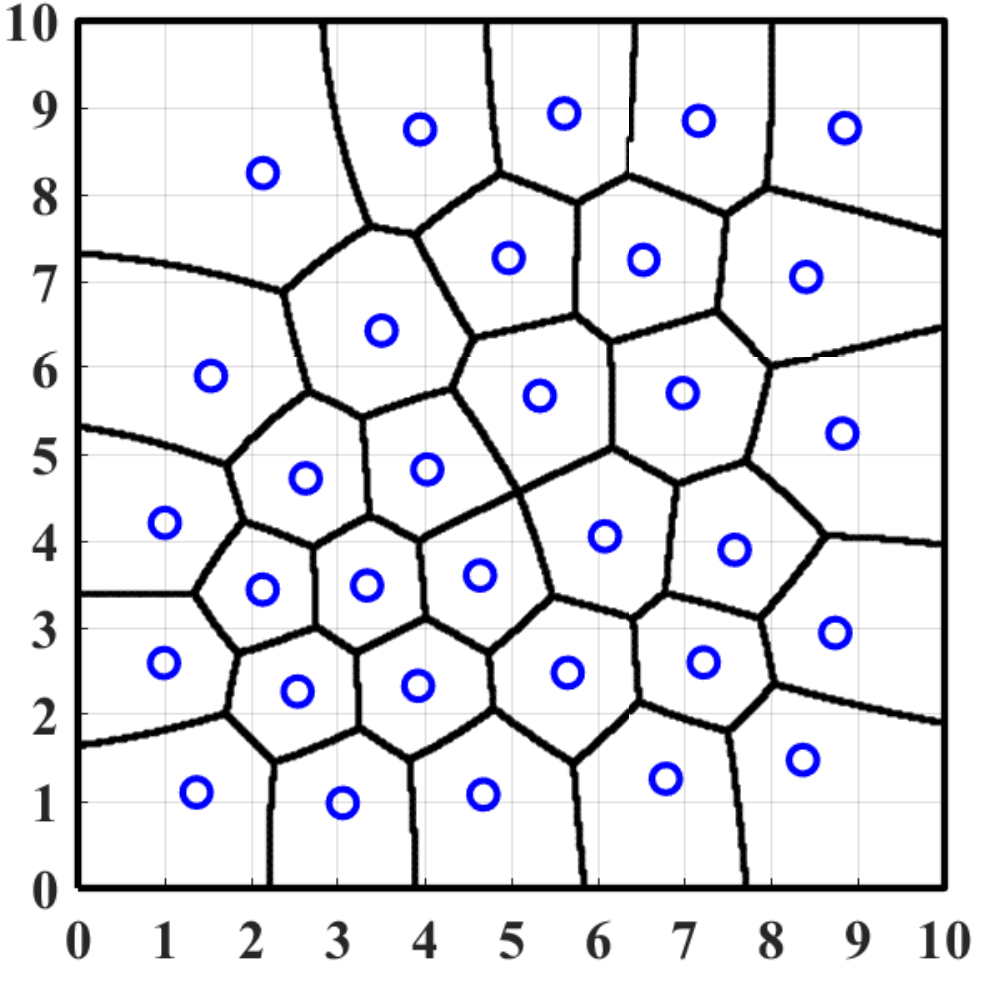}
\label{nonuniformPartitions32}}
\hfil
\subfloat[]{\includegraphics[width=2.8in]{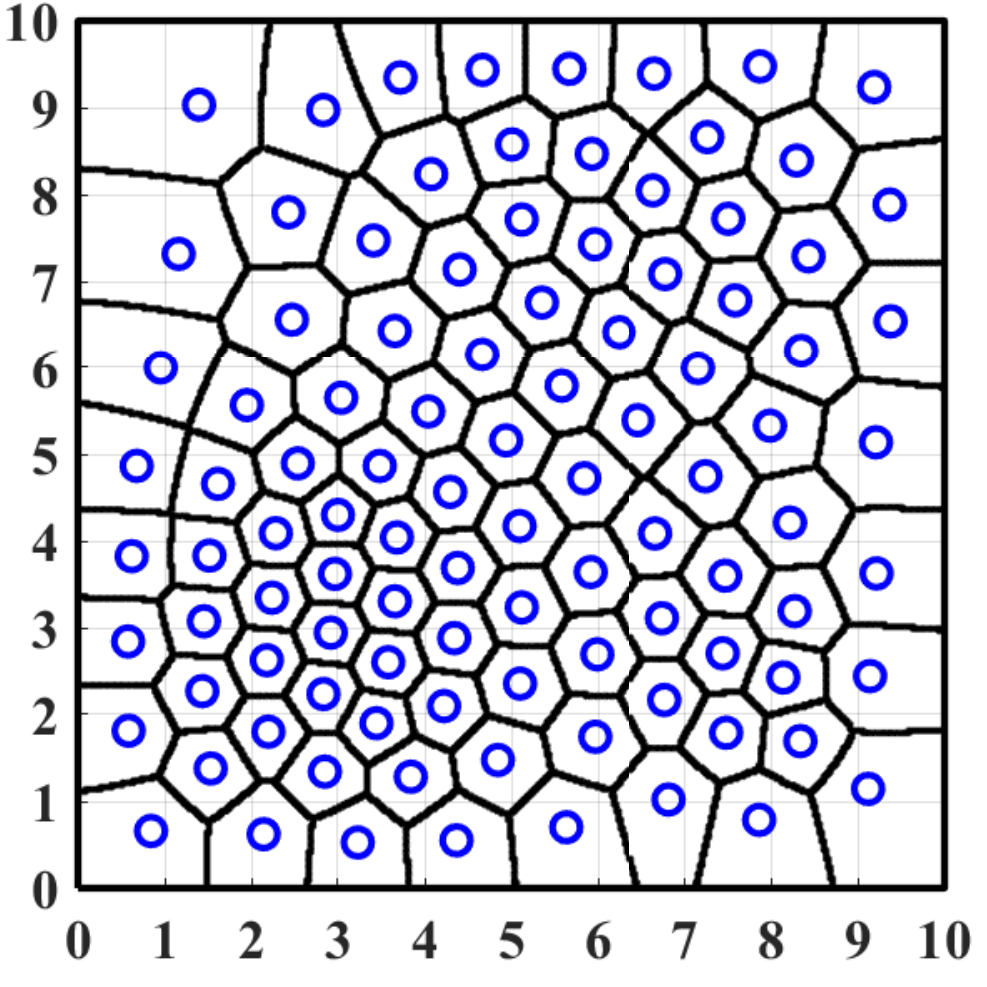}
\label{nonuniformPartitions100}}
\captionsetup{justification=justified}
\vspace{-2ex}
\caption{\small{The UAV projections on the ground with generalized Voronoi Diagrams where $\alpha=2$ and the source distribution is non-uniform. 
(a) 32 UAVs. (b) 100 UAVs.}}
\label{uniformDistortionPartition3}
\end{figure}
\else   \fi \section{Conclusion}
We studied quantizers with parameterized distortion measures for an application to UAV deployments.  Instead of using the
traditional mean distance square as the distortion, we introduce a distortion function which models the energy
consumption of UAVs in dependence of their heights.  We derived the unique parameter optimized quantizer -- a uniform
scalar quantizer with an optimal common parameter -- for uniform source density in one-dimensional space.  In addition, two
Lloyd-like algorithms are designed to minimize the distortion in two-dimensional space.  Numerical simulations
demonstrate that the common weight property extends to two-dimensional space for a uniform density.

\ifarxiv
\appendices
\section{Proof of \lemref{lem:ggam}}\label{app:proof_lemma_ggam}
  To find the optimal $1-$level parameter quantizer $(x^*,h^*)$ for a uniform density $\lam(\ome)=1/A$, we need to
  satisfy \eqref{eq:criticalpoint}, i.e., for\footnote{Note, there is no optimizing over the regions, since there is only
  one.} $\Ome=\Vor_1=\Vor_1^*=[0,A]$
  \begin{align}
    0&=\int_0^A (x^*-\ome)\big( (x^*-\ome)^2+h^{*2}\big)^{\gam-1}  d\ome.
  \end{align}
  Substituting $x^*-\ome$ by $\ome$ we get 
  \begin{align}
    0&=\int_{x^*-A}^{x^*} \ome \big( \ome^2+h^{*2}\big)^{\gam-1}  d\ome.
  \end{align}
  Since the integral kernel is an odd function in $\ome$ and $x^*\in[0,A]$, it must hold
  \begin{align}
    0=-\int_0^{x^*-A} \ome(\ome^2+h^{*2})^{\gam-1}d\ome + \int_0^{x^*}\ome(\ome^2+h^{*2})^{\gam-1}d\ome
    \intertext{by substituting $\ome$ by $-\ome$ we get}
    \int_{0}^{A-x^*} \ome(\ome^2+h^{*2})^{\gam-1}d\ome = \int_0^{x^*} \ome(\ome^2+h^{*2})^{\gam-1}d\ome.
  \end{align}  
  Hence for any choice of $h^*$ it must hold $x^*=A-x^*$, which is equivalent to $x^*=A/2$.
  To find the optimal parameter, we can just insert $x^*$ into the average distortion
  \begin{align}
    \AvDis(x^*,h) &= \frac{1}{A}\int_0^A \frac{(x^*-\ome)^2+h^2)^{\gam} }{h}d\ome
    = \frac{1}{A} \int_0^{A/2} \frac{(\ome^2+h^2)^{\gam}}{h}d\ome\label{eq:AvDisxstar}
  \intertext{where we substituted again and inserted $x^*=A/2$. By substituting $\ome$ with $2\ome/A$ 
    and $h$ with $u=2h/A$  we get}
    &= \int_0^1  \frac{2}{A}\frac{((A\ome/2)^2 +   (Au/2)^2)^{\gam}}{u}d\ome
      =\left(\frac{A}{2}\right)^{2\gam-1} \int_0^1 f(\ome,u,\gam)d\ome
  \end{align}
  where for each $\gam\geq 1$ the integral kernel $f$ is a convex function in $\vx=(\ome,u)$ over $\R_+^2$. Let us
  rewrite $f$ as
  \begin{align}
    f(\ome,u,\gam)= \frac{(\ome^2+u^2)^{\gam}}{u}= \frac{\Norm{(\ome,u)}_2^{2\gam}}{u}.
  \end{align}
  Clearly, $\Norm{\vx}_2$ is a convex and continuous function in $\vx$ over $\R^2$ and since $(\cdot)^{2\gam}$ with
  $2\gam\geq2$ is a strictly increasing continuous function, the concatenation $f(\vx,\gam)$ is a strict convex and
  continuous function over $\R_+^2$. Hence, for any $\vx_1,\vx_2\in\R^2$ we have
  \begin{align}
    \Norm{\lam\vx_1+(1-\lam)\vx_2}_2^{2\gam}<\lam \Norm{\vx_1}_2^{2\gam} + (1-\lam)\Norm{\vx_2}_2^{2\gam}
  \end{align}
  for all $\lam\in(0,1)$. But then we have also for any $u_1,u_2\in\R^2_+$ and $\ome\geq 0$ 
  \begin{align}
    f(\lam u_1 +(1-\lam)u_2,\ome,\gam) < 
    \frac{\lam\Norm{(\ome,u_1)}_2^{2\gam} + (1-\lam)\Norm{(\ome,u_2)}_2^{2\gam}}{\lam u_1+(1-\lam)u_2}
    \label{eq:fnormgam}.
  \end{align}
  Considering the following inequality 
  \begin{align}
    \frac{1}{u_1}\! +\!\frac{1}{u_2} &=
       \left(\frac{1}{u_1}\! + \!\frac{1}{u_2}\right)\frac{\lam u_1 \!+\!(1\!-\!\lam)u_2}{\lam u_1 \!+\! (1\!-\!\lam)u_2}
       =\frac{ \left(\lam\! +\!\frac{(1\!-\!\lam)u_2}{u_1}\! +\! (1\!-\!\lam) \!+\! \frac{\lam u_1}{u_2}\right)}{\lam
       u_1 \!+\! (1\!-\!\lam)u_2} 
       > \frac{1}{\lam u_1\! +\! (1\!-\!\lam)u_2}\notag
  \end{align}
  and \eqref{eq:fnormgam}, we will have
  \begin{align}
    f(\lam u_1 +(1-\lam)u_2,\ome,\gam) < \lam f(u_1,\ome,\gam) + (1-\lam)f(u_2,\ome,\gam) 
  \end{align}
  for every $\lam\in(0,1)$. Hence, the integral kernel is a strictly convex function for every $\ome\geq0,\gam\geq 1$, and since the
  infinite sum (integral) of convex functions is again a convex function, for $u>0$, we have shown convexity of $F(u,\gam)$. 
  Note, $f(u,\ome,\gam)$ is continuous in $\R_+^2$ since it is a product of the continuous functions
  $\Norm{(u,\ome)}_2^{2\gam}$ and $1/(u+0\cdot\ome)$, and so is $F(u,\gam)$. 
  Therefore, the only critical point of $F(\cdot,\gam)$ will be the unique global minimizer
  \begin{align}
    g(\gam)=\arg\min_{u>0} F(u,\gam),
  \end{align}
  which is defined by the  vanishing of the first derivative:
  \begin{align}
    F'(u)  &\!= \!\int_0^{1}\! (\ome^2\!+\!u^2)^{\gam-1} \left( (2\gam\!-\!1)-\frac{\ome^2}{u^2}\right)d\ome
    \!=\!\frac{1}{u^{2}}\!\int_{0}^{1} (\ome^2\!+\!u^2)^{\gam-1}\left((2\gam\!-\!1)u^2-\ome^2\right)d\ome\label{eq:Fderivative}.
  \end{align}
  Hence, $F'(u)$ can only vanish if $u<1/\sqrt{2\gam-1}$, which is an upper bound on $g(\gam)$.
  The optimal parameter for minimizing the average distortion  \eqref{eq:AvDisxstar} is then 
  \begin{align}
    h^*= \frac{A}{2} g(\gam) \quad\text{with}\quad \AvDis(x^*,h^*)= \left(\frac{A}{2}\right)^{2\gam-1} g(\gam).
  \end{align}
  Analytical solutions for $F'(u)=0$ are possible for integer valued $\gam$.  Let us set $0<x=u^2$ in
  \eqref{eq:Fderivative}, then for $\gam\in\N$, the integrand in \eqref{eq:Fderivative} will be a polynomial in $\ome$
  of degree $2\gam$ and in $x$ of degree $\gam$. For $\gam\in\{1,2,3\}$ the integrand will be 
  \begin{align}
    (\ome^2+x)^0 (1x-\ome^2)&=x-\ome^2\\
    (\ome^2+x)^1 (3x-\ome^2)&=3x^2+2\ome^2x -\ome^4 \\
    (\ome^2+x)^2 (5x-\ome^2)&=5x^3 +9\ome^2 x^2 +3\ome^4 x -\ome^6
  \end{align}
  which yield with the definite integrals to
  \begin{align}
    0 &= \ome(x-\frac{\ome^2}{3})\Big|_{\ome=1}\label{eq:xfirst}\\
    0 &= \ome( 3x^2 +\frac{2\ome^2x}{3} -\frac{\ome^4}{5}  )\Big|_{\ome=1}\label{eq:xtwo}\\ 
    0 &= \ome( 5x^3 + 3\ome^2x^2 +\frac{3\ome^4x}{5}  -\frac{\ome^6}{7}  )\Big|_{\ome=1}\label{eq:ccubic} 
  \end{align}
  Solving \eqref{eq:xfirst} for $x$  yields to the only feasible solution
  \begin{align}
    x=\frac{1}{3} \quad\RA\quad g(1)=\frac{1}{\sqrt{3}}\approx 0.577.
  \end{align}
  The solutions of \eqref{eq:xtwo} are 
  \begin{align}
    x_{\pm}= -\frac{1}{9} \pm \sqrt{\frac{1}{81}+\frac{1}{15}} = \frac{\pm \sqrt{32/5} -1}{9}
  \end{align}
  Since only positive roots are allowed, we get as the only feasible solution
  \begin{align}
    g(3)=\frac{\sqrt{\sqrt{32/5}-1}}{3}\approx 0.412.
  \end{align}
  Finally, the cubic equation \eqref{eq:ccubic} results in
  \begin{align}
    5x^3 + 3 x^2 + \frac{3}{5} x - \frac{1}{7}=0
  \end{align}
  The solution of a cubic equation can be found in \cite[2.3.2]{Zwi03} by calculating the discriminant
  \begin{align}
    \Del=q^2+4p^3 \quad\text{with}\quad q=\frac{2b^3-9abc+27a^2d}{27a^3},p=\frac{3ac-b^2}{9a^2}
  \end{align}
  Let us identify $a=5,b=3,c=3/5$ and $d=-1/7$, then we get
  \begin{align}
    q&=\frac{6\cdot 9-9\cdot 9 - 27\cdot 5^2\cdot1/7}{27\cdot 5^3}
    =-\frac{3}{3\cdot 5\cdot 25} -\frac{1}{5\cdot 7}=-\frac{32}{25\cdot35}\\
    \Del&=q^2 + 4 \left(\frac{3\cdot 3 -9}{9\cdot 5^2}\right)^3=q^2>0
  \end{align}
  which indicates only one real-valued root, given by
  \begin{align}
    x=\alp_+^{1/3}+\alp_-^{1/3}-\frac{b}{3a} 
    \quad\text{with}\quad  \alp_{\pm}=\frac{-q\pm\sqrt{\Del}}{2} = \left\{0, \frac{32}{25\cdot 35} \right\}
  \end{align}
  which computes to
  \begin{align}
    x=\left( \frac{32}{5^3\cdot 7}\right)^{1/3}- \frac{1}{5}
    =\frac{(\frac{32}{7})^{1/3}-1}{5} \RA g(5)=\sqrt{\frac{(\frac{32}{7})^{1/3}-1}{5}} \approx 0.363.
  \end{align}
\fi 

\ifarxiv%
\section{Proof of \lemref{lemma:allActive}}\label{app:proof_lemma_active}
  Although, this statement seems to be trivial, it is not straight forward to show.
  We will use the quantization relaxation for the average distortion $\AvDis$ in \eqref{eq:Pbar} to  show that the
  $N-$level parameter optimized quantizer has strictly smaller distortion than the $(N-1)-$level optimized quantizer
  \eqref{eq:optPbar}. We define, as in quantization theory, see for example \cite{GN98}, an \emph{$N-$level quantizer}
  for $\Ome$, by a (disjoint) partition $\Rset=\{\Rset_n\}_{n=1}^N\subset \Ome$ of $\Ome$ and assign to each partition
  region $\Rset_n$ a quantization-parameter point $(\vp_n,h_n)\in\Ome\times\R_+$. The assignment rule or
  \emph{quantization rule} can be anything such that the regions are independent of the value of the quantization and
  parameter points.  Minimizing over the quantizer, that is, over all partitions and possible quantization-parameter
  points will yield to the parameter optimized quantizer, which is by definition the optimal deployment  which generate
  the generalized Voronoi regions as the optimal partition (tessellation\footnote{Since we take here the continuous
  case, the integral will not distinguish between open or closed sets.}). This holds for any density function
  $\lam(\ome)$ and target area $\Ome$.  To see this\footnote{We use the same argumentation as in the prove of
    \cite[Prop.1]{KJ17}.}, let us start with any quantizer $(\vP,\vh,\Rset)$ for $\Ome$ yielding to the average
    distortion 
  \begin{align}
    \AvDis(\vP,\vh,\Rset) &=\sum_{n=1}^N \int_{\Rset_n} \Dis(\vp_n,h_n,\ome)\lam(\ome) d\ome \geq \sum_{n=1}^N \int_{\Rset_n} 
    \left(\min_{m\in[N]} \Dis(\vp_m,h_n,\ome) \right) \lam(\ome)d\ome\notag\\
    &=\int_{\Ome} \min_{m\in[N]} \Dis(\vp_m,h_n,\ome) \lam(\ome)d\ome
    =\sum_{n=1}^{N} \int_{\Vor_n(\vP,\vh)} \Dis(\vp_n,h_n,\ome) \lam(\ome)d\ome
  \end{align} 
  where the first inequality is only achieved if for any $\ome\in \Rset_n$ we have chosen $(\vp_n,h_n)$ to be the
  optimal quantization point with respect to $\Dis$, or vice versa, if every $(\vp,h_n)$ is optimal for every $\ome\in
  \Rset_n$, which is the definition of the generalized Voronoi region $\Vor(\vP,\vh)$. Therefore, minimizing over all
  partitions gives equality, i.e.
  \begin{align}
    \min_{\Rset} \AvDis(\vP,\vh,\Rset)=\AvDis(\vP,\vh,\Vor(\vP,\vh))
  \end{align}
  for any $(\vP,\vh)\in\Ome^N\times\R_+^N$.
  Hence, we have shown that the parameterized distortion quantizer optimization problem is equivalent to the locational-parameter
  optimization problem
  \begin{align}
    \min_{\vP\in\Ome^N,\vh\in\R_+^N} \min_{\Rset\in\Ome^N} \AvDis(\vP,\vh,\Rset) 
    = \min_{\vP\in\Ome^N,\vh\in\R_+^N} \AvDis(\vP,\vh,\Vor(\vP,\vh))=\AvDis(\vP^*,\vh^*,\Vor^*).\label{eq:optquanteqoptdeploy2}
  \end{align}
  We need to show that for the optimal $N-$level parameter-quantizer $(\vP^*,\vh^*,\Vor^*)$ with $\Vor^*=\Vor(\vP^*,\vh^*)$,
  we have $\mu(\Vor_n)>0$ for all $n\in[N]$.  Let us first show that each region is indeed a closed interval, i.e.,
  $\Vor_n^*=[b^*_{n-1},b^*_n]$ with $0\leq b^*_{n-1}\leq b^*_n\leq A$. 
  
  By the definition of the Möbius regions in \lemref{lem:moebiusdia}, each dominance region is either a single interval
  (if it is a ball not contained in the target region or a halfspace) or two disjoint intervals (if its a ball contained
  in the target region), we can not have more than $K_n\leq 2N-2$ disjoint closed intervals for each Möbius (generalized
  Voronoi) region.  Therefore, the $n$th optimal Möbius region is given as $\Vor^*_n=\bigcup_{k=1}^{K_n}v_{n,k}$, where
  $v_{n,k}=[a_{n,k-1},a_{n,k}]$ are intervals for some $0\leq a_{n,k-1}\leq a_{n,k}\leq A$.

  Let us assume there are quantization points with disconnected regions, i.e. $K_n>1$ for $n\in\Ind_d$ and some
  $\Ind_d\subset[N]$. Then, we will re-arrange the partition $\Vor^*$ by concatenating the $K_n$ disconnected
  intervals $v_{n,k}$ to $\Rset_n=[b_{n-1},b_{n}]$ for $n\in\Ind_d$ and move the connected regions appropriatly such that
  for all $n\in[N]$ it holds $\mu(\Rset_n)=\mu(\Vor^*_n)=b_n-b_{n-1}$ and $b_{n-1}\leq b_{n}$, where we set $b_0=0$ and
  $b_N=A$. For the new concatenated regions, we move each $q_n^*$ to the center of the new arranged regions, i.e.,
  $\tq_n=\frac{b_n+b_{n-1}}{2}$ for $n\in\Ind_d$. If for the connected regions $n\in[N]\setminus\Ind_d$, the
  quantization point $q^*_n$ is not centroidal, by placing it at the center of the corresponding closed interval,
  we will obtain a strictly smaller distortion by \lemref{lem:ggam}. Hence, for the optimal quantizer, the quantization
  points must be centroidal and we can assume $\tq_n=(b_n+b_{n-1})/2$ for all $n\in[N]$. 
  In this rearrangement, we did not change the parameters $h_n^*$ at all.  The rearranged partition
  $\Rset=\{\Rset_n\}$ and replaced quantization points $\tvp=(\tq_1,\dots,\tq_N)$ provide the average distortion 
  \begin{align}
    \AvDis(\tvp,\vh^*,\Rset)=\sum_{n=1}^N \int_{b_{n-1}}^{b_n} \frac{ ((\tq_n\!-\!\ome)^2+h_n^{*2})^{\gam}}{h_n^{*}} d\ome
      = 2\sum_{n=1}^N \int_{0}^{\frac{b_n-b_{n-1}}{2}} \frac{ (\ome^2+h_n^{*2})^{\gam}}{h_n^{*}} d\ome
  \end{align}
  where we substituted $\ome$ by $\tq_n\!-\!\ome$. Since the function
  $(\ome^2+h_n^{*2})^{\gam}$ is strictly monotone increasing in $\ome$ for each $\gam>0$, for any $n\in\Ind_d$, we have
  \begin{align}
  \AvDis_n= \AvDis(\tq_n,h_n^*,\Rset_n)=    2\int_{0}^{\frac{b_n-b_{n-1}}{2}} \frac{ (\ome^2+h_n^{*2})^{\gam}}{h_n^{*}} d\ome 
    < \sum_{k=1}^{K_n}\int_{a_{n,k}-q_n^*}^{a_{n,k-1}-q_n^*} \frac{ (\ome^2+h_n^{*2})^{\gam}}{h_n^{*}}
    d\ome\label{eq:Avdisn}
  \end{align}
  since the non-zero gaps in $\bigcup_k [a_{n,k}-q_n^*,a_{n,k-1}-q_n^*]$ will lead to larger $\ome$ in the RHS integral
  and therefore to a strictly larger average distortion.  Therefore, the points $(\tvp,\vh^*)$ with closed intervals
  $\{\Rset_n\}$ have a strictly smaller average distortion, which contradicts the assumption that $(\vp^*,\vh^*)$ is the
  parameter-optimized quantizer \eqref{eq:optquanteqoptdeploy}. Hence, $K_n=1$ for each $n\in[N]$ and every $\gam\geq 1$.
  Moreover, the optimal quantization points must be centroids of the intervals, i.e. $x_n^*=(b_n^*+b^*_{n-1})/2$.


  Now, we have to show that the optimal quantization regions $\Vor_n^*=\{[b^*_{n-1},b^*_n]\}_{n=1}^N$ are not points, i.e.,
  it should hold $b^*_n>b^*_{n-1}$ for each $n\in[N]$. If $b^*_n=b^*_{n-1}$ for some $n$, then the $n$th average distortion $\AvDis_n$
  will be zero for this quantization point, since the integral is vanishing. But, then we only optimize over $N-1$
  quantization points. So we only need to show that an additional quantization point strictly decreases the minimum
  average distortion.  Hence, take any non-zero optimal quantization region $\Vor_n^*=[b^*_{n-1},b^*_n]$. We know by
  \lemref{lem:ggam} that the optimal quantizer $q_n^*$ for some closed interval $\Vor_n^*$  must be centroidal for
  any parameter $h_n$.  Hence, if we split  $\Vor_n^*$ with  $\mu_n^*=b^*_n-b^*_{n-1}$ by a half and put two quantizers
  $q_{n_1}$ and $q_{n_2}$ with the same parameter $h_n^*$ in the center, we will get by using \eqref{eq:Avdisn}
  \begin{align}
      \AvDis_{n_1}\!+\!\AvDis_{n_2}&=\frac{1}{h_n^*} 
     \left( \int_{b^*_{n-1}}^{b^*_{n-1}+\frac{\mu_n^*}{2}} ( (q_{n_1}-\ome)^2 + h_n^{*2})^{\gam}d\ome
     +\int_{b_{n-1}^*+\frac{\mu_n^*}{2}}^{b_n^*}((q_{n_2}-\ome)^2+h_n^{*2})^{\gam}d\ome\right)\notag\\
   \intertext{Substituting $q_{n_i}-\ome$ by $\ome$, we get} 
     & =  \int_{-\frac{\mu_n^*}{4}}^{\frac{\mu_n^*}{4}} \frac{(\ome^2+h_n^{*2})^{\gam}}{h_n^*}d\ome
     + \int_{-\frac{\mu_n^*}{4}}^{\frac{\mu_n^*}{4}} \frac{(\ome^2+h_n^{*2})^{\gam}}{h_n^*}d\ome\\
     & =  2\int_{0}^{\frac{\mu_n^*}{4}} \frac{(\ome^2+h_n^{*2})^{\gam}}{h_n^*}d\ome 
     + w\int_{0}^{\frac{\mu_n^*}{4}}    \frac{ (\ome^2+h_n^{*2})^{\gam}}{h_n^*}d\ome\\
     & <  2\!\int_{0}^{\frac{\mu_n^*}{4}} \frac{(\ome^2\!+\!h_n^{*2})^{\gam}}{h_n^*}d\ome
     + 2\!\int_{\frac{\mu_n^*}{4}}^{\frac{\mu_n^*}{2}} \frac{(\ome^2\!+\!h_n^{*2})^{\gam}}{h_n^*}d\ome
     = 2\!\int_0^{\frac{\mu_n^*}{2}} \frac{(\ome^2\!+\!h_n^{*2})^\gam}{h_n^*} d\ome=\AvDis_n.
  \end{align}
  Hence, the average distortion will strictly decrease if $\mu_n^*>0$. Therefore, the $N-$level parameter optimized
  quantizer will have quantization boundaries $b_n\!>\!b_{n\!-\!1}$ for $n\in[N]$.%
\fi 
\ifarxiv

\section{Proof of \thmref{thm:commonheight}}\label{sec:proof_theorem} 
  We know by \lemref{lemma:allActive} that the optimal quantization regions are closed non-vanishing
  intervals $\Vor_n^*=[b_{n-1}^*,b_n^*]$ for some $b^*_{n-1}<b_n^*$ with quantization points
  \begin{align}
   \vp_n^*= x_n^*=\frac{b^*_n+b^*_{n-1}}{2}\label{eq:xnoptimal}
  \end{align}
  for $n\in[N]$.
  Let us set $\mu_n^*=b_n^*-b_{n-1}^*$ for $n\in[N]$. By substituting $\frac{2(x_n^*-\ome)}{\mu_n}=\tome$ and
  $\hGlob_n=\frac{u^*_n \mu_n^*}{2}$  in the average distortion, we get 
  \begin{align}
    \AvDis(\vP^*,\vh^*,\Vor^*)&=\sum_{n=1}^N \int_{b_{n-1}^*}^{b_n^*} 
       \frac{ ((\pGlob_n-\ome)^2+{\hGlob_n}^2)^{\gam}}{\hGlob_n}\frac{d\ome}{A} 
       = \sum_{n=1}^N \int_{1}^{-1} -\frac{  (\mu_n^{*2} \tome^2/4 + u_n^{*2}\mu_n^{*2}/4)^{\gam}}{u_n^* \mu_n^*/2}
    \frac{\mu_n}{2A}d\tome\notag\\
    &= \frac{1}{ 2^{2\gam-1}A}\sum_{n=1}^N \mu_n^{*2\gam}\cdot\int_0^1 \frac{(\ome^2+u^{*2}_n)^{\gam}}{u^*_n}d
    \ome\label{eq:AvDismutilde} 
  \end{align}
  where we used \eqref{eq:xnoptimal} to get for the integral boundaries
  $2(\pGlob_n-b_{n-1}^*)/\mu_n^*=1=-2(\pGlob_n-b_{n}^*)/\mu^*_n$.  We do not know the value of $u^*_n$ and $\mu_n^*$ but
  we know that $\mu_n^*>0$ and $\sum_{n=1}^N\mu_n^*=A$ by \lemref{lemma:allActive}. Furthermore, \eqref{eq:AvDismutilde} is
  the minimum over all such  $\mu_n>0$ and $u_n>0$. Hence, it must hold
  \begin{align}
    \AvDis(\vP^*,\vh^*,\Vor^*)
    = \frac{1}{2^{2\gam-1}A} \min_{u_n>0} \min_{\substack{\mu_n>0\\ A\!=\! \sum_{n\!=\!1}^N \mu_n}} 
    \sum_{n=1}^N \mu_n^{2\gam}\cdot\left(\int_0^1 
      \frac{(\ome^2+u^2_n)^{\gam}}{u_n}d \ome\right)
      = \frac{g(\gam)}{2^{2\gam-1}A}  
         \min_{\substack{\mu_n>0\\ A\!=\! \sum_{n\!=\!1}^N \mu_n}} \sum_{n=1}^N \mu_n^{2\gam}  \notag
  \end{align}
  where in the last equality we used \lemref{lem:ggam}.  By the Hölder inequality we get for $p=2\gam,
  q=2\gam/(2\gam-1)$ 
  \begin{align}
    \sum_{n=1}^N \mu_n^{2\gam}=\sum_{n=1}^N \mu_n^p  = \sum_{n=1}^N \mu_n^p \cdot\Big(\sum_{n=1}^N (1/N)^q \Big)^{p/q} \cdot N
    \geq\Big( \sum_{n=1}^N \frac{\mu_n }{N}\Big)^p \cdot N =
    \left(\frac{A}{N}\right)^{2\gam}N\notag
  \end{align}
  where the equality is achieved if and only if $\mu_n^*=A/N$. Hence, the optimal parameter-quantizer is the 
  uniform scalar quantizer $x_n^*=(2n-1)A/2N$ with identical parameters $h^*=(A/2N)g(\gam)$ resulting in the
  minimum average distortion \eqref{eq:optimumavpow}.

  Let us note here, that for identical parameters, the Möbius regions are closed intervals and reduce to Euclidean
  Voronoi regions by \lemref{lem:moebiusdia},
  for which the optimal tessellation is known to be the uniform scalar quantizer, see for example \cite{GN98}. 
\fi

\section*{References} 
\printbibliography
\end{document}